\documentclass[a4paper,american]{scrartcl}

\usepackage{lmodern}
\usepackage[T1]{fontenc}
\usepackage[utf8]{inputenc}
\usepackage[american]{babel}
\usepackage{latexsym}
\usepackage{amsmath}
\usepackage{amssymb}
\usepackage{braket}
\usepackage{graphicx}
\usepackage{verbatim}

\usepackage[authoryear,round]{natbib}
\usepackage[indentfirst=false,font=small]{quoting}
\usepackage[pdfusetitle, bookmarks=false, breaklinks=false,pdfborder={0 0 0},backref=false,colorlinks=false]{hyperref}

\renewcommand{\vec}{\boldsymbol}

\begin{document}

\title{The Wave-Function as a Multi-Field}

\author{Mario Hubert\thanks{Université de Lausanne, Faculté des lettres, Section de philosophie, 1015 Lausanne, Switzerland. E-mail: \href{mailto:Mario.Hubert@unil.ch}{Mario.Hubert@unil.ch}}         \and
Davide Romano\thanks{E-mail: \href{mailto:davideromano1984@libero.it}{davideromano1984@libero.it}}
}

\maketitle
 {\centering \emph{Forthcoming in the European Journal for Philosophy of Science} \par}
 
\begin{abstract}
It is generally argued that if the wave-function in the de Broglie--Bohm theory is a physical field, it must be a field in configuration space. Nevertheless, it is possible to interpret the wave-function as a multi-field in three-dimensional space. This approach hasn't received the attention yet it really deserves. The aim of this paper is threefold: first, we show that the wave-function is naturally and straightforwardly construed as a multi-field; second, we show why this interpretation is superior to other interpretations discussed in the literature; third, we clarify common misconceptions. 
\end{abstract}

\tableofcontents
\newpage
\section{Can the Wave-Function Be a Field?}

The wave-function is a peculiar object: it is mathematically defined in configuration space, and yet it is supposed to determine the motion of particles in three dimensions. We need to translate this mathematical picture into a coherent ontological story. Since the particles and the wave-function are defined in too different mathematical spaces, it has been generally agreed that the wave-function cannot be interpreted as a field in three-dimensional space. Whereas the electromagnetic field assigns to every point in three-dimensional space a unique field value, the wave-function doesn't do so.

One strategy to save a field interpretation would be to declare configuration space as the fundamental space of the world. In this space, the wave-function indeed assigns to every point a unique value. The following passage by John Bell is usually interpreted to suggest such that:
\begin{quotation}
No one can understand [the de Broglie--Bohm] theory until he is willing to think of $\psi$ as a real objective field rather than just a `probability amplitude'. Even though it propagates not in $3$-space but in $3N$-space. \citeyearpar[][p.\ 128]{Bell:1987aa}
\end{quotation} 
The most developed interpretation along these lines is the marvelous point interpretation by David \citet{Albert:1994aa}. 

Another strategy for a field interpretation of the wave-function was pursued by Travis \citet{Norsen:2010aa}. In fact, this is rather a new theory than just a mere interpretation because the wave-function is reduced to local fields in three-dimensional space governed by a revised Schrödinger equation. 

In this paper, we steer a middle course between these two proposals. We aim at showing that the wave-function can be indeed construed as a new kind of field in three-dimensional space, namely, as a \emph{multi-field}. We will defend this interpretation against Albert's interpretation and Norsen's theory. For the multi-field incorporates the best of both worlds: an ontology in three-dimensional space without changing the mathematical formalism. 

\section{The Multi-Field}
\label{sec:multi-field}

The multi-field view starts from the idea to generalize a classical field, which specifies a definite field value for each location of three-dimensional space. A charged particle that is posited at a given location will feel the force generated by the value of the field at this location. The multi-field generalizes this concept to $N$-tuples. Given an $N$-particle system, a multi-field specifies a precise value for the entire $N$-tuple of points in three-dimensional space, thus determining, given the actual positions of $N$ particles, the motion of all particles.

We suggest that the wave-function is the mathematical representation of such a multi-field. In the first-order formulation of the de Broglie--Bohm theory, the multi-field specifies the velocity of the particles according to the guiding equation, whereas in the second-order formulation it specifies the acceleration of the particles, according to the quantum Newtonian law: $F_C + F_Q = ma$. In other words, the multi-field assigns a ``multi-velocity''  to the configuration of particles in the first-order theory  and a ``multi-acceleration'' induced by a multi-quantum-force to the configuration of particles in the second-order theory. The multi-velocity and multi-acceleration then generate the velocities and accelerations for each single particle of the configuration.  

There is an important difference between the classical field and the multi-field. In the classical case, the field of $N$ particles can always be decomposed into a sum of the single-particle fields, because every particle produces its own field. As the wave-function is not produced by ``quantum sources'', the multi-field of an $N$-particle system is not decomposable into a sum of single-particle quantum fields. If the wave-function is factorizable, it may be decomposable into single-particle wave-functions, but each of these wave-functions is not generated by its corresponding particle. Thus, the multi-field is rather a holistic or relational field assigned to sets of $N$ particles. 

The idea of the multi-field is not novel. It dates back to \citet[Ch.\ 5]{Forrest:1988aa}, where he introduced the concept of \emph{polywaves} as a generalization of classical \emph{mono-waves}: 

\begin{quotation}
I posit polywaves, which are disturbances to polyadic fields. The familiar monowaves (monadic waves) are assignments to each location of some member of the set of possible field-values for that location. Likewise an $[N]$-adic polywave is an assignment to each ordered n-tuple of locations of a member of the set of possible field-values for that $[N]$-tuple of locations. The integer $[N]$ is just the ``number of particles''. And the possible field-values are $[N]$-adic relations. \citeyearpar[p.\ 155]{Forrest:1988aa}
\end{quotation}

In the original proposal by Forrest, the concept of polywaves is defined in the framework of standard quantum mechanics. The problem with this approach is that the integer $N$ in the definition above cannot be straightforwardly understood as the number of particles, since in standard quantum mechanics a system is described only by the wave-function. This is the reason why Forrest's metaphysics seems to be constituted by pure relations between (empty) points of space, which seems to be a rather peculiar ontology. This problem is solved in the de Broglie--Bohm theory, where the integer $N$ refers to the number of real particles, and the relations between particles are easily explained by the dynamical correlations induced by the multi-field as we have described above.

\citet{Belot:2012aa} was then the first to apply Forrest's polywaves to the de Broglie--Bohm theory and to dub this approach the \emph{multi-field interpretation.} After a brief sketch of the multi-field view, Belot dismisses it for the following four reasons (pp. 72-3):
\begin{enumerate}
\item The multi-field doesn't have sources.
\item The multi-field violates the action-reaction principle.
\item The multi-field doesn't restore energy-momentum conservation.
\item The multi-field transforms under Galilean boosts differently from the electromagnetic field. 
\end{enumerate}
Objections 1, 2, and 4 arise from the requirement that the multi-field must have essentially the same features as classical fields. We expect the multi-field to have different physical features than its classical counterpart. This is not problematic for the existence of the multi-field, but also desirable: if a multi-field behaved like a classical field, there would be no point of introducing a novel entity in the ontology of the physical world. 

A crucial question is to understand which features are essential to fields in general and which are only essential to classical fields. We want to make this distinction starting from a definition of a classical field, and then giving a generalization of it for the multi-field.\footnote{However, we will not enter here in the metaphysical issue concerning the nature of fields. As the notion of a classical field can be framed in different metaphysical views (for example, Humean view, dispositional view, etc.), the same procedure is in principle applicable to the multi-field.} We can think of a classical field to be defined by the following features:
\begin{enumerate}
\item[(a)]
it is an assignment of intrinsic properties to the points of space it is defined on, and
\item[(b)]
it ensures energy and momentum conservation.
\end{enumerate}
Now, in the case of the multi-field, we must substitute (a) with
\begin{enumerate}
\item[(c)]
it is an assignment of intrinsic properties to particular $N$-tuples of points of three-dimensional space.
\end{enumerate}
In sum, we suggest that the definition of a multi-field is captured by statements (b) and (c), and that only classical fields are required to obey (a) and (b). So, energy conservation is an important factor to reify a mathematical function into a physical field, and we will see indeed that the multi-field permits to restore energy-momentum conservation in the de Broglie-Bohm theory. Now, we are ready to answer Belot's objections against the multi-field.

\subsection*{No Sources and no Action-Reaction}
We think that these two features are intrinsically connected with each other: a particle must act back on a field if it has generated the field itself. In the de Broglie-Bohm theory, the multi-field is not produced by particles, and therefore it is plausible that the action-reaction principle is violated. However, this is not problematic for the theory (since it is physically coherent) nor for considering the wave function as a field, since the action-reaction principle is not included in the defining statements (b) and (c).

\subsection*{Energy-Momentum Conservation}
In the de Broglie-Bohm theory, there is energy-momentum conservation for closed systems, and this is assured by the wave-function. This can be easily shown within the Hamilton-Jacobi formulation of the theory, where the classical potential $V$ and the quantum potential $Q=-\frac{\hbar^2}{2m}\frac{\nabla^{2}{\Psi}}{\Psi}$ together contribute to energy-momentum conservation. In particular, if $V$ and $Q$ are time-independent (that is, if the system is closed), the total energy of particles along their trajectories is conserved \citep[][p.\ 285]{Holland:1993aa}:
$$
\sum_{i=1}^{N}\frac{1}{2}mv_i^2+V+Q=\text{const}.
$$
A similar relation holds for momentum conservation, in the absence of external (classical and quantum) forces. If $\sum_{i=1}^{N}\nabla_i(V+Q)=0$ then
$$ 
\sum_{i=1}^{N}\vec{p}_i=\text{const.}
$$
A concrete example may help to illustrate the situation. Consider a one-particle system in an infinite potential well. The wave-function of the system inside the well is $$\Psi(x,t)=A\sin(kx)e^{\frac{i}{\hbar}Et},$$ with total energy $$E_{tot}=\frac{k^{2}\hbar^{2}}{2m}.$$
Yet, the kinetic energy of the particle is zero ($v=\frac{\partial{S}}{\partial{x}}=0$, since the wave function is real), and the classical potential energy $V$ is zero (by definition inside the potential well). So where does the energy of the system come from? It comes from the quantum potential. For we have $$E_{tot}=E_{\text{kin}}+V+Q=0+0-\frac{\hbar^2}{2m}\frac{\nabla^{2}{\Psi}}{\Psi}=\frac{k^{2}\hbar^{2}}{2m}.$$ In sum, the total energy of the system is absorbed by the quantum potential, which is produced by the wave-function. Interpreting the wave-function as a multi-field permits to account for energy conservation in a natural way.

\subsection*{Galilean Transformation}
Regarding the transformation properties of the multi-field, it is a mathematical fact that the wave-function transforms differently under Galilean boosts than the electromagnetic field. The correct transformation involves a multiplication of the boosted wave-function by a plane wave with the same velocity. We can accept this transformation behavior of the multi-field as a non-classical feature arising from a non-classical law, the Schrödinger equation. Furthermore, a plane wave can be split up into two parts, one dependent on space and the other independent of space. The space-independent part can be interpreted as a phase of the wave-function. But this phase factor doesn't change the motion of particles. Therefore, this would indicate that the mathematical representation of the multi-field is just determined up to a phase factor. Moreover, this objection, like the one of the action-reaction principle, does not undermine the possibility to regard the wave-function as a field, since the ordinary Galilean transformation is not included in the statements (b) and (c).

\bigskip

We think that the the multi-field account naturally explains the non-local behavior of Bohmian particles: since the value of the multi-field depends on $N$-tuples of points and not on single points, the behavior of a given configuration of particles is intrinsically non-local, as it were a single structure moving in three-dimensional space. The multi-field is thus a new type of field mathematically represented by the wave-function. It fills the physical space with precise values for every $N$-tuple of points. Particles will feel a certain velocity and a certain acceleration depending on the actual configuration $(\vec{x}_1,\dots,\vec{x}_N)$ and depending on the precise value of the quantum field $\Psi(\vec{x}_1,\dots,\vec{x}_N)$.

Although the multi-field is a physical field in three-dimensional space, its mathematical representation is given by the usual wave-function in configuration space.\footnote{\citet{Chen:2017ab} challenges the view of defining the wave-function in mathematical terms; he instead proposes a nominalistic approach.} Therefore, the multi-field interpretation does not have to be confused with Norsen's theory of local fields, which aims at defining the wave-function itself in three dimensions. We show how this idea differs from the multi-field in the next section.

\section{A Multitude of Fields Is Not a Multi-Field}
\label{sec:multitude}
Travis \citet{Norsen:2010aa} proposed a Bohmian quantum theory, in which there is no longer a wave-function in configuration space \citep[see also][]{Norsen:2015ab}. Instead, the main dynamical entity is the conditional wave-function associated to every particle in three-dimensional space.\footnote{The conditional wave-function $\psi_t(x)$ of a particle is defined by the universal wave-function $\Psi$, once the positions of all the other particles in the universe $Y(t)$ are fixed: $\psi_t(x):=\Psi(x,Y(t))$.} The one-particle conditional wave-functions, however, don't suffice to recover all the predictions of the de Broglie--Bohm theory, since they cannot describe entangled states between particles. Norsen presents a nice example. 

\begin{figure}[ht]
\centering
\includegraphics[width=7cm]{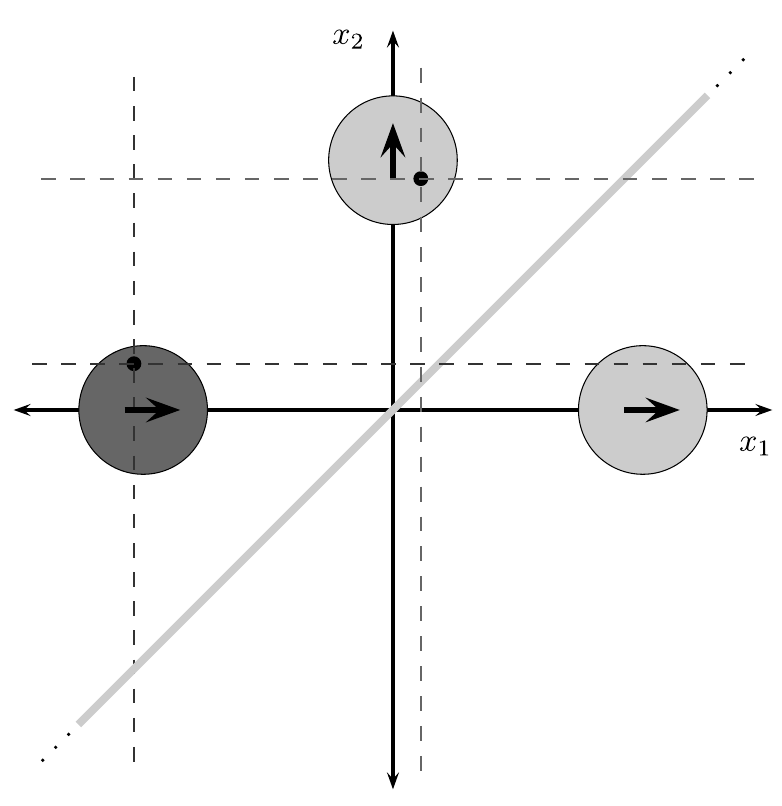}
\caption{Scattering of two non-entangled particles represented in the two--particle configuration space. A particle moves parallel to the $x_1$-axis approaching a particle that sits at $x_2=0$. The potential of the resting particle is marked as a light grey diagonal line. Their initial wave-function is marked in dark grey. After  scattering, the first moving particle stops, and the other particle moves upwards. The wave-function is then in a superposition indicated by two light grey wave-functions. \citep[Picture from][p.\ 1867]{Norsen:2010aa}}
\label{fig:conditional-wf-1}
\end{figure}

\begin{figure}[ht]
\centering
\includegraphics[width=7cm]{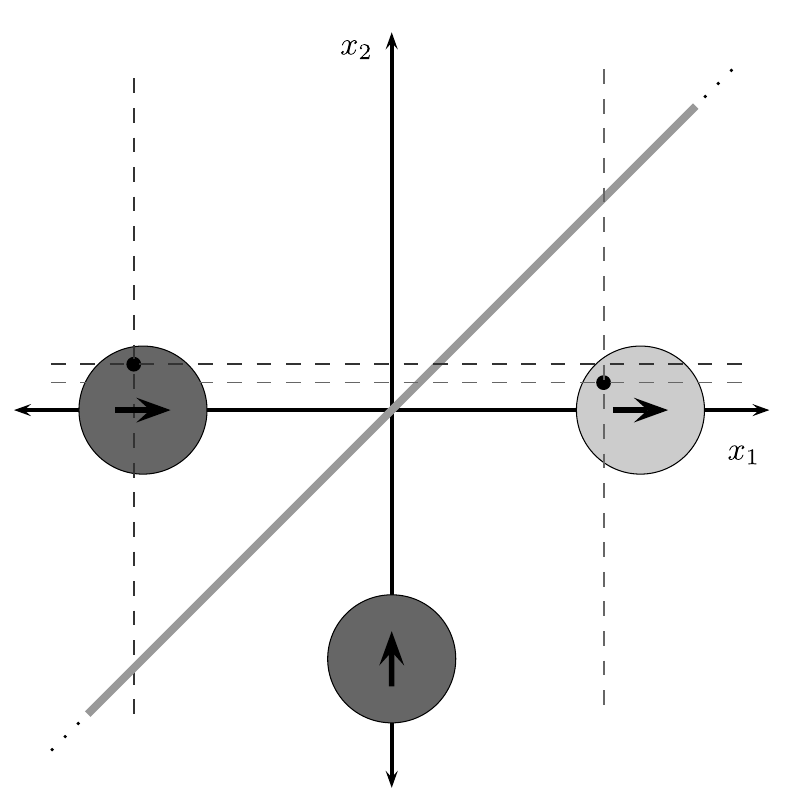}
\caption{Scattering of two entangled particles represented in the two--particle configuration space. As in Fig.\ \ref{fig:conditional-wf-1}, a particle approaches a resting particle from the left and collides at $x_2=0$. Their entangled initial wave-function is depicted in dark grey. After scattering, the resting particle starts moving to the right parallel to the $x_1$-axis, while the other particle stops at $x_2=0$. The post--scattering wave-function is drawn in light grey. \citep[Picture from][p.\ 1868]{Norsen:2010aa}}
\label{fig:conditional-wf-2}
\end{figure}

Imagine two particles that are about to collide. We can prepare the system in two different ways. In the first case, we start with a non-entangled wave-function (see Fig.\ \ref{fig:conditional-wf-1}), and in the second case, we prepare the system to be entangled (see Fig.\ \ref{fig:conditional-wf-2}). The initial particle positions are the same. And, more importantly, the initial conditional wave-functions of both particles are the same, too. Yet, we can prepare each system in such a way that the particles move differently after collision. 

This shows that conditional wave-functions cannot do the job alone in retrieving all Bohmian trajectories. While conditional wave-functions can render the correct trajectories in the first example, 
they cannot do so for the entangled state. The information about entanglement gets lost in the definition of conditional wave-functions---that's the same for the reduced density matrix in an EPR experiment, where it merely gives us the statistics for one particle irrespective of what happens to the other particle.

Norsen's idea is therefore to add additional local fields to the conditional wave-functions to recover quantum entanglement. The task of these new fields is to change the conditional wave-functions of each particle in such a way that they render the correct trajectories even if the system is entangled; in fact, these fields are non-zero only if there is entanglement.

Norsen's theory of exclusively local beables makes the very same empirical predictions as the de Broglie--Bohm theory---it even predicts the very same trajectories---, but the price to be paid is a more contrived law for the evolution of all those local fields. First, it turns out that there are infinitely many such interacting fields since the evolution of the interaction fields requires further interaction fields\dots a never ending recursion. And it's not clear yet that one can get satisfactory results with only a finite set of these fields. Second, each conditional wave-function follows a modified Schrödinger equation, in which the other interaction fields are included. And these interaction fields themselves have their own evolution equation. This makes the theory mathematically very complicated and almost impractical for calculating empirical predictions.

We now realize that Norsen \emph{does not} present a theory of a multi-field. There are infinitely many ordinary local fields in physical space, which interact. In fact, this is instead a theory of a multitude of ordinary fields!

Still, Norsen's theory of local beables is sometimes confused with the multi-field:
\begin{quotation}
We may consider the multi-field option---this postulates a multitude of fields in
3d space, corresponding to each physical particle. Each of these fields is determined by
the wavefunction in configuration space, [\dots]
Yet, at each
instant in time, a field is defined in 3-D space corresponding to each particle. \citep[][p.\ 3211]{Suarez:2015aa}
\end{quotation}
It is apparent that this is not the multi-field account that we have presented in the previous section. Indeed, in the multi-field account, there is no ``multitude of fields'' on physical space; rather, there is just one field that assigns a value to a set of $N$ particles. You cannot decompose these values corresponding to each fields assigned to each particle because, for entangled systems, one cannot attribute a wave-function to a single particle. 

\section{Against a Field in Configuration Space}
\label{sec:configuration-space}

David \citet{Albert:1996aa} argued that if the wave-function is a field it has to be a field in configuration space. In doing so, he takes this space as the fundamental space of physics, in which only one single universal particle exists:

\begin{quote}
On Bohm's theory, for example, the world will consist of exactly two physical objects. One of those is the universal wave-function and the other is the universal \emph{particle}.

And the story of the world consists, in its entirety, of a continuous succession of changes of the \emph{shape} of the former and a continuous succession of changes in the \emph{position} of the latter.

And the dynamical laws that \emph{govern} all those changes -- that is: the Schrödinger equation and the Bohmian guidance condition -- are completely deterministic, and (in the high-dimensional space in which these objects live) completely \emph{local}. \citep[][p.\ 278]{Albert:1996aa}
\end{quote}

The major reason for Albert to develop an ontology in configuration space is to have the wave-function as a \emph{local beable}: the wave-function is determined by the local values it assigns to every point of configuration space. Hence, the motion of the universal particle is completely determined by the field value at its location, exactly as in classical electrodynamics, where the motion of a charged particle is determined by the value of the electromagnetic field at its location. We doubt, however, that this kind of locality is a universal principle for physical theories, especially in the quantum domain.\footnote{Many arguments were given to criticize Albert's ontology \citep[for instance,][]{Chen:2017aa,Maudlin:2013aa,Monton:2013aa}. We will focus on one argument that we think is the strongest with respect to the multi-field interpretation.}

In our opinion, the major drawback of this interpretation is that it doesn't distinguish between the mathematical space and physical space. The fact that a physical object is mathematically defined in configuration space, does not necessarily imply that the object itself has to exist in configuration space. Nor does it imply that the world is $3N$ dimensional. We encounter a similar issue in the meaning of dimensions in classical mechanics.
The number of dimensions to mathematically describe a physical object is given by the number of degrees of freedom that we need to describe the state and the motion of that
object. For example, a rigid body that translates and rotates is represented in a six-dimensional space. But, of course, this does not mean that the rigid body is really a six-dimensional object: in fact, it exists in a three-dimensional space, but we need six degrees of freedom to fully describe its state of motion. Here, the distinction between the mathematical representation and the actual physics is obvious, since everybody agrees on the ontology, on what a rigid body is. 

Albert takes a different approach in quantum mechanics, as well as in classical mechanics. He starts with the mathematical formalism and searches for criteria in how to distill an ontology. His primary principle is locality: find the mathematical space in which the physical entities are locally defined. You may also call this principle \emph{separability}, since it is about the ontology of objects and not about their dynamical behavior.\footnote{More explicitly, there are two types of locality: \emph{ontological} locality and \emph{dynamical} locality. The former coincides with separability and is about local beables, while the latter is about the \emph{behavior} of physical objects, for instance, Bell's notion of \emph{local causality} or Einstein's locality.} For the de Broglie--Bohm theory this principle distinguishes configuration space as the space in which the wave-function is locally defined. So, he takes this space as fundamental. 

In classical mechanics, the ontology in three-dimensional space and in phase space are both separable. So here the locality principle does not single out one space over the other. Therefore, Albert invokes a second principle: take the local space that has the lowest dimension. Applying this principle, we get that the fundamental space of classical mechanics is three dimensional.

We think that Albert's approach to extract an ontology from the mathematical formalism of physics relies on a heterodox understanding of what physical theories are. Physical theories are not bare mathematics \citep[for this critique, see also][]{Maudlin:2013aa}. Only one part consists of formulas; the other part consists of a commentary to the formulas in order to tell us what these symbols refer to \emph{in the real world}. This distinguishes physics from mathematics. In particular, the standard commentary of the formalism of the de Broglie--Bohm theory relates the mathematics to $N$-particles in three-dimensional space. We agree that mathematical equations do not uniquely determine what they refer to, but we disagree that the principle of separability singles out a preferred ontology.  

The primary task of physics is to explain what we observe, to explain our manifest image \citep[to use an expression of][]{Sellars:1963aa}. A theory does so by proposing a scientific image, namely an ontology with corresponding laws of nature.  We think that Albert's separability criterion may lead to  ontologies that hardly fill the gap between the scientific and manifest image. Albert's scientific image consists of a point in configuration space that is locally guided by the wave-function. He gets the manifest image by a functional analysis of the Hamiltonian in the Schrödinger equation. The Hamiltonian has the structure of giving rise to a three-dimensional world, with tables and chairs, on a coarse-grained level. 

In the multi-field view, the scientific image consists of many particles in three dimensions, guided by a non-local field in this very space. The advantage of this view is that the scientific image and the manifest image are situated in the very same space. So the macroscopic objects in the manifest image are mereologically composed of microscopic particles from the scientific image. This mereological composition is lacking in Albert's ontology. Although he introduces particles in three-dimensions, they are mere \emph{shadows} of the marvelous point \citep[][p.\ 130]{Albert:2015aa}.  
Indeed, it is not sufficient to explain the manifest image just by the dynamical laws but also by how these laws give rise to the manifest image from the fundamental ontology. 
That real particles exist in
three-dimensional space composing tables and chairs is much more plausible and straightforward than a marvelous point in a very high dimensional space explaining the behavior of tables and chairs by a functional analysis of the Hamiltonian \citep[see also][for an argument along these lines]{Emery:2017aa}. We think that we should prefer ontologies that are closer to the manifest image of the physical world over those representing a significant departure from it.

\section{Advantages} 

In the following, we list the advantages of the multi-field interpretation in more systematic order. Some of the points have been mentioned above, but here we make them explicit. 

\subsection*{General Physical Interpretation}

The multi-field interpretation is largely independent of the concrete formulation of the de Broglie--Bohm theory in terms of a first-order formulation \citep{Durr:2013fk,Valentini:2010aa}, second-order formulation \citep{Bohm:1993aa}, or quantum Hamilton-Jacobi formulation \citep{Holland:1993aa}. In the first-order formulation the multi-field specifies the velocities of each particle. In the second-order formulation, the acceleration is specified by means of the quantum potential. We think, however, that the multi-field interpretation is particularly useful for the second-order theory, because it allows to retrieve the entire classical scheme of how motion is generated: field $\rightarrow$ potential $\rightarrow$ force $\rightarrow$ acceleration. 

Recent interpretations, like the nomological interpretation \citep{Goldstein:2013ab}, the Humean interpretation \citep{Bhogal:2015aa,Callender:2015aa,Esfeld:2012mz,Miller:2013aa}, or the dispositional interpretation \citep{Esfeld:2012mz,Suarez:2015aa} require a commitment to what laws of nature are. The multi-field interpretation, on the other hand, shows that one can have an ontological interpretation of the wave-function without a commitment to the status of laws of nature. We think  that the wave-function is best regarded as an objective physical field than a nomological entity, since the wave function is in general time-dependent.  Even if the universal wave-function turned out to be time-independent, it will still be a solution of a dynamical equation, like the Wheeler-DeWitt equation. Having a law for a nomological entity seems to us not in the spirit of a nomological entity. This is different for  the Hamiltonian in classical mechanics, which does not arise from a law, and therefore may be interpreted as a nomological entity. These two arguments suggests to regard the wave-function as a field, similarly to the electromagnetic field (being a solution of the Maxwell equations).

\subsection*{The Entire Ontology in Three Dimensions}
That Bohmian particles are guided by the wave-function is often taken merely as a heuristic  metaphor. As the wave is defined in configuration space, it cannot directly influence the motion of particles. Having the wave-function as a multi-field, however, gives this intuition an ontological underpinning. The de Broglie--Bohm theory is hence a pilot-multi-wave theory, where the wave directly guides particles in three-dimensional space. This has also the advantage that there is an ontological continuity between one-particle and many-particle scenarios. A one-particle wave-function cannot be only visualized in three-dimensional space because this space mathematically coincides with configuration space, but there is indeed a wave in three-dimensional space. When this one-particle wave-function gets entangled with an external $N$-particle system, the new $(N+1)$-particle system is still guided by a wave in three dimensions. The double-slit experiment provides a vivid example of the explanatory advantage of this view. While the particle goes through one of the slit, the wave literally enters both slits, thereby determining the motion of the particle and accounting for the characteristic interference pattern on the screen.

\subsection*{Simplicity}

The multi-field view is different from Norsen's approach, which reduces the universal wave-function to a set of one-particle wave-functions. There, each particle has an associated local “pilot-wave” in three-dimensional space, so that an $N$-particle wave-function reduces to a set of local beables. But in order to make the theory empirically adequate and to account for entanglement, Norsen has to introduce an infinite number of interacting fields for the guiding fields. Thus, the formalism of the theory becomes extraordinarily complicated. This is due to the recursive structure of the local Schrödinger equation, leading to a kind of Taylor series for every local guiding field comprising a particle. This would be a totally new structure for a fundamental law of nature. Since the total infinite series of fields cannot be calculated, it is an open question at what order one can truncated this series to have approximately good empirical results. We think that Norsen’s attempt to write down a theory of exclusively local beables shows how unnatural it is to mathematically embed the wave-function in three-dimensional space. 

The multi-field view, on the other hand, is much simpler: it is an interpretation of the wave-function as a new type of field in three-dimensional space. In particular, it does not require us to modify the definition of the wave-function, and so it does not require to modify the mathematical formalism of the theory. Quantum non-locality is encoded by having the mathematics in an abstract high-dimensional space. This explains quantum non-locality in the simplest way.

\subsection*{Instantiating a Non-Local Beable}

For \citet{Maudlin:2013aa}, the wave-function in the de Broglie--Bohm theory refers to a real physical entity that determines the behavior of particles. The wave-function is according to him best regarded as the mathematical representation of a new physical object: the \emph{quantum state}. 
Yet, contrary to Albert, Maudlin suggests that we should not extract the ontology of the quantum state directly from its mathematical representation for the following reasons:
\begin{enumerate}
\item The wave-function contains some degrees of freedom that are merely gauge, since they do not lead to different quantum states. One example is the overall phase: wave-functions with different overall phases represent the same quantum state, since their empirical predictions are exactly the same. 
\item The wave-function (for an $N$-particle system) is naturally expressed in configuration space because it refers to the positions of a real configuration of $N$ particles in three-dimensional physical space. 
\end{enumerate}

Nevertheless, \citeauthor{Maudlin:2013aa} does not specify which sort of physical entity the quantum state is. He merely describes it as a \emph{non-local beable}:
\begin{quote}
From the magisterial perspective of fundamental metaphysics, then, our precise quantum theories have a tripartite ontology: a space-time structure that assumes a familiar approximate form at mesoscopic scale; some sort of local beables (particles, fields, matter densities, strings, flashes) in that space-time; and  a single universal non-local beable represented by the universal wave-function $\Psi$. \citep[p. 356]{Maudlin:2015aa}
\end{quote}

The multi-field idea, on the other hand, goes beyond Maudlin's characterization of the wave-function: it is a genuine physical field. Because of its properties, the multi-field is a non-local beable. It is a beable since it is a real physical object. And it is non-local since the value of the multi-field is specified not for one point but only for $N$-tuples of points in three-dimensional space. The multi-field thus explains what the quantum state is, and it instantiates a non-local beable. This relation of instantiation between a non-local beable and the multi-field could be also analyzed in terms of grounding in the sense of \citet{Schaffer:2009aa}. A thorough discussion of this issue, however, would go beyond the scope of this paper.

\section{Objections and Replies}

In the course of writing the paper, we have received some critical remarks about the multi-field approach. We now mention the most important ones followed by our reply.

\begin{enumerate}
\item \emph{The multi-field doesn't give us a new ontological understanding of the wave-function.}

We regard the multi-field interpretation as a new ontological interpretation of the wave-function that has been so far ignored. This approach is incompatible with the nomological interpretation of the wave-function, and its metaphysical characterizations in terms of dispositionalism and Humeanism. But one may still ground the multi-field on an extended Humean mosaic that also comprises contingent non-reducible relations \citep{Darby:2012aa}. In this way, we would have a Humean interpretation, in which the wave-function is no longer a nomological entity because it is not part of the best system. The mosaic would be comprised of points composing three dimensional space, a particle configuration, and the multi-field. Similarly, \citet{Loewer:1996aa} embedded Albert's marvelous point interpretation in a Humean framework. The difference is, however, that Loewer's mosaic is separable in configuration space, while Darby's mosaic is non-separable in three-dimensional space.

\item \emph{You don't change the formalism. How can the wave-function be something in three-dimensional space, if it is defined on configuration space?}

There is a difference between the mathematical structure that we use to define a physical object and the ontology of this object. The multi-field interpretation is based on this distinction. The configuration space is the mathematical space that we need to describe a function which generally depends on $3N$ degrees of freedom (where $N$ is the number of particles of the system); three-dimensional space is the physical space in which the object represented by that function is defined.  

A simple example can be of some help here. Historically, the first formulation of classical mechanics was due to Newton.  Newton's theory of mechanics was an ontological theory, that is, a theory with clear ontological commitments: particles moving in three-dimensional space accelerated by force acting on them. The same theory can be casted in the Hamiltonian formulation. Here, the system is represented by a particle moving in phase space with a trajectory described by the Hamiltonian function. Nevertheless, it is understood that the Hamiltonian formulation is just a mathematical representation of the ontological picture of classical mechanics given by Newton's theory. What we usually do in practice is to use both formulations simultaneously: the physical ontology of Newton's theory (that's what the world is built of) and the mathematics of the Hamiltonian formulation (which is often more convenient in doing calculations).

Quantum mechanics has followed an inverse path: the formalism of the theory is an extension of the Hamiltonian formulation of classical mechanics. And the historical mistake was to try to extract from this formulation the physical interpretation of the theory. We regard the de Broglie--Bohm theory as the ontological theory of quantum mechanics (the analogue of Newton's theory for the classical case), which is a theory of particles moving in three-dimensional physical space acted upon by a multi-field. In a nutshell, the multi-field interpretation bears the advantages of Norsen's ontology (that is, having the wave-function physically as a field on three-dimensional space) and the simplicity of the standard formalism (that is, the wave-function as a mathematical object in configuration space). 

\item \emph{The multi-field runs into the problem of communication.}

\citeauthor{Suarez:2015aa} claims the following:

\begin{quote}
According to this view, the multi-fields are defined at each instant by the wave-function in configuration space, and the question is how the wave-function “communicates” to physical three-dimensional space in order to fix each of the fields and the positions of the particles for any system of N particles. […] Also, note that the communication is curiously one way: while the wave-function fixes the physical properties, including positions, of particles in three-dimensional space, these have no effect back onto the wave-function, which essentially ignores which are the actual particle trajectories amongst all the possible trajectories compatible with the dynamics. \citeyearpar[p.\ 3212]{Suarez:2015aa}
\end{quote}

It is difficult to see how the problem of communication arises in the multi-field approach. Indeed, there is a problem of communication when different physical entities live in \emph{different} physical spaces and yet influence each other. But, in the case of the multi-field approach, both the guiding wave and the particles live in the very same space, namely, three-dimensional space. So, there is no problem of communication in the multi-field account. 

Moreover, the fact that the communication is “one way” is independent of the problem of communication; it ought to be an objection to regarding the wave-function as a field in the first place. This can become a problem if we want the wave-function to behave like a classical field, where particles and field mutually affect each other. But there are good reasons to think that, if the wave-function is a field, it must certainly be a new type of field, possessing completely new physical features. The action-reaction principle may thus be thought of as a characteristics of classical fields, and it is better to abandon this principle in the ontology of quantum theory. After all, there is no logical inconsistency in thinking of particles not acting back on the field. This could just confirm that the type of physical interaction between the (Bohmian) particle and the multi-field is after all not a classical interaction. 

\item \emph{Fields have to be local and propagate with finite speed.}

We disagree that these features are essential to fields, although they are crucial to classical fields. In section \ref{sec:multi-field}, we proposed that the multi-field is specified by the statements (b) and (c), that is, by an assignment of properties to $N$-tuples of points of three-dimensional space and momentum-energy conservation for closed systems. All other features about locality and  dynamics are derived from the equations that implement these fields in the theory. 

\item \emph{Physicists have always understood the wave-function as a kind of multi-field.}

We could not find historical evidence that the multi-field used to be the default view among physicists, not even the default view among Bohmians. Since de Broglie's and Bohm's work, physicists were struggling to understand the ontological status of the wave-function because it is mathematically defined in configuration space, and yet it communicates with particle configurations in three-dimensional space. This tension has lead people to claim that we live in configuration space, or that we must give up the idea that the wave-function literally guides the particles. We think that both scenarios are problematic. We argue, instead, that the wave-function can exist as a certain kind of physical field on three-dimensional space, its mathematical definition notwithstanding. 

\item \emph{How can the multi-field fill up physical space? For example, if I point to a location of physical space, what part of the multi-field is there?}

The multi-field is a non-local beable; therefore, it does not assign any value to single points of space according to Bell's definition of local beables. The multi-field exists as a physical field in three-dimensional space by assigning properties to all $N$-tuples of points of space. The ontology of particles, on the other hand, is local because particles occupy single points in space. The connection between the non-local beable and the local beables is accomplished by the guidance equation, which projects the non-local property assigned to the $N$-tuple into local properties of particles. 

One may still object that if the multi-field doesn't assign a value to any point of three-dimensional space, the multi-field cannot exist on this space.  But this objection presupposes that the multi-field is a local beable. The multi-field exists at each point of space as a a non-separable or relational field, which needs $N$ spatial points to deliver a field value. Without specifying these $N$ points the multi-field does not give you a precise value. The field value assigned to an $N$-tuple has metaphysical priority, since this property cannot be built up from local properties. Nevertheless, for an \emph{actual} configuration of particles the multi-field gives every particle a local property in form of a specific velocity (first-order formulation) or a specific acceleration (second-order formulation).

Non-local beables are unfamiliar in physics and apart from the wave-function hard to find. But one can define a non-local quantity in General Relativity that shares similar features with the multi-field. This quantity goes by the name of \emph{ADM energy} (Arnowitt, Deser, and Misner). In General Relativity, it's not possible to define a local energy density for the gravitational field, unless one invokes additional (unjustified) structures on space-time, like a preferred coordinate system  \citep[][section 11.2]{Wald:1984aa}. Still, it's possible to define a global energy on an asymptotically flat space-times by means of an integral over an infinitely large 2-sphere. This integral is the ADM energy. In this case, too, it would be meaningless to ask what the energy is at a single point of space-time, since the energy is only globally defined. There are no local values for the ADM energy, like there are no local values for the multi-field. 

\end{enumerate}

\section{Conclusion}

We have argued that the multi-field interpretation has been dismissed for the wrong reasons owing to prejudices from classical fields. Construing the Bohmian wave-function as a multi-field is actually the most conservative physical interpretation compared to Albert's marvelous point interpretation and Norsen's local fields theory. For it describes the entire ontology of the theory in three-dimensions without changing the mathematical formalism, thus solving the problem of communication and the problem of perception, and it provides a natural metaphysical explanation for the non-local behavior of particles. 

Tim Maudlin regards the wave-function to be a non-local beable building on the work of John Bell. The multi-field interpretation starts from a completely different route, namely, by an analysis and extension of the classical field concept. In our view, the wave-function is also a non-local beable because it is, indeed, a multi-field.

\section*{Acknowledgements}
We wish to thank David Albert, Guido Bacciagaluppi, Michael Esfeld, Dustin Lazarovici, Tim Maudlin, Matteo Morganti, Travis Norsen, Andrea Oldofredi, Charles Sebens, and Tiziano Ferrando for many helpful comments on previous drafts of this paper. We also thank the audience of the 3\textsuperscript{rd} Annual Conference of the Society for the Metaphysics of Science (SMS) and especially Lucas Dunlap for commenting on our paper at this event. We also thank two anonymous referees for their very detailed reviews. Davide Romano's research was funded by the Swiss National Science Foundation (grant no.\ 105212\_149650).

\bibliographystyle{abbrvnat}
\bibliography{references}

\end{document}